\documentclass[12pt]{article}
\usepackage{amssymb,amsmath,latexsym}
\newtheorem{thm}{Theorem}[section]
\newtheorem{cor}[thm]{Corollary}
\newtheorem{prop}[thm]{Proposition}

\newtheorem{rem}[thm]{Remark}
\newtheorem{dfn}[thm]{Definition}
\newtheorem{ex}[thm]{Example}
\numberwithin{equation}{section}
\newenvironment{proof}{\trivlist \item[\hskip
	\labelsep{{\it\underline{Proof}.\/}}]}{\ \rule{0.5em}{0.5em}}%
{\endtrivlist}
{\endtrivlist}
\begin{document}
\title{\bf Leibniz bialgebras, classical Yang-Baxter equations and dynamical systems}
\author{A. Rezaei-Aghdam$^{a}$\thanks{e-mail: rezaei-a@azaruniv.ac.ir}\,\,\, {\tiny }L. Sedghi-Ghadim$^{b}$\thanks{e-mail: sedghi88@azaruniv.ac.ir and leila.sedghi@gmail.com} and GH. Haghighatdoost$^{b}$\thanks{e-mail: gorbanali@azaruniv.ac.ir}\,\,\\ \\
	{\small {$^a${\em Department of Physics, Azarbaijan Shahid Madani
				University , 51745-406, Tabriz, Iran.}}}\\
	{\small {$^b${\em Department of Mathematics, Azarbaijan Shahid Madani
				University , 51745-406, Tabriz, Iran.}}}}
\maketitle
\hspace{13.5pt}
\begin{abstract}
This work is intended as an attempt to extend the notion of bialgebra for Lie algebras to Leibniz algebras and also, the correspondence between the Leibniz bialgebras and its dual is investigated. Moreover, the coboundary Leibniz bialgebras, the classical $r$-matrices and Yang-Baxter equations related to the Leibniz algebras are defined, and some examples are given. Finally, a method for construction of a dynamical system on a Leibniz manifold via Leibniz bialgebra is presented.
\end{abstract}
\textbf{Mathematics Subject Classifications ($2010$):} 17A32, 17B62, 16T25.

\section{Introduction}
\hspace{13.5pt}
The notion of Leibniz algebras was first proposed by Blokh \cite{Blo} in 1965 under the name $D$-algebras as a natural generalization of Lie algebras. Leibniz algebras later were rediscovered by J. L. Loday \cite{Lod1} who called them Leibniz algebras as noncommutative analogues of Lie algebras. Leibniz algebras are defined by a bilinear bracket  satisfying the derivation identity which is non-antisymmetric. In the past two decades the theory of Leibniz algebras has been extensively studied
and many results of Lie algebras have been extended to Leibniz algebras, such as, classical results on Cartan subalgebras \cite{Cas,Alb1,Omir1,Ayup1}. Up until now the Levi's theorem for Leibniz algebras \cite{Bar}, the representation, homology and cohomology of Leibniz algebras \cite{Lod2,Cuv&Pira} (see also \cite{Lod3}) also, the classification of low dimensional solvable and nilpotent Leibniz algebras \cite{Ayup2} has been studied in this manner. However, the structure of the theory of Leibniz algebras mostly remains unexplored. Here, we will define bialgebra structure related to the Leibniz algebras; as an extension of Lie bialgebras \cite{Drin} (see for a review \cite{Sch}). Our method is in the definition of Leibniz bialgebra based on cohomological view, which is essentially different from the method used in \cite{Eli} that based on algebraic view. Recently, this concept has been studied in \cite{Sheng} using the quadratic Leibniz algebra. One can see the importance of Leibniz algebra due to applications in mathematical physics, such as in \cite{Sheng2,Kotov}.\newline
The paper is organized as follows. In Section $2$, we recall main definitions  and results about Leibniz algebras
and their cohomology. Then, along the lines of Ref. \cite{Sch}, in Section $3$ we introduce the notion of Leibniz bialgebras and show that if $(\mathfrak{g},\gamma)$ be
a Leibniz bialgebra then its dual $(\mathfrak{g}^{\ast},\mu^{t})$ will be also a Leibniz bialgebra where $\mu$ is a Leibniz bracket on $\mathfrak{g}$. The definitions of coboundary Leibniz bialgebra, classical $r$-matrices and classical Yang-Baxter equation are presented in Section $4$.  Section $5$ provides a detailed exposition of a computation method for finding examples of Leibniz bialgebras and also  the related classical $r$-matrices where some two and three dimensional examples are given. Finally, in Section $5$ we present a method for construction of a dynamical system on a Leibniz manifold using Leibniz bialgebra. Concluding remarks and open problems are discussed in the conclusion section.
\section{Preliminaries}
 Let us recall some basic definitions about Leibniz algebras.
\begin{dfn}\cite{Lod3}
A right or left Leibniz algebra $\mathfrak{g}$ is a vector space over a field $\mathbb{F}$ endowed with a bilinear bracket $[., .]$ satisfying the following right or left Leibniz identity, respectively
\begin{align}
&[[Y,Z],X]=[[Y,X],Z]+[Y,[Z,X]],\quad \forall X,Y,Z\in \mathfrak{g},\label{RLI}\\
	\text{or}\cr
&[X,[Y,Z]]=[[X,Y],Z]+[Y,[X,Z]],\quad \forall X,Y,Z\in \mathfrak{g}.\label{LLI}
\end{align}
For any $X\in \mathfrak{g}$,  consider the right (left) adjoint mapping $\mathrm{ad}_{X}^{(r)}:\mathfrak{g}\longrightarrow \mathfrak{g}$ $(\mathrm{ad}_{X}^{(l)}:\mathfrak{g}\longrightarrow \mathfrak{g})$ defined by $\mathrm{ad}_{X}^{(r)}(Z)=[Z,X]\,(\mathrm{ad}_{X}^{(l)}(Z)=[X,Z])$. Clearly, the right (left) Leibniz identity is equivalent to assert that $\mathrm{ad}_{X}^{(r)}\,(\mathrm{ad}_{X}^{(l)})$ is a derivation for any $X\in \mathfrak{g}$\footnote{Note that, contrary to the Lie algebra the bracket of Leibniz algebra is not antisymmetric i.e. $[X,Y]\neq -[Y,X]$. This is the main difference between Leibniz algebra and Lie algebra.}.
\end{dfn}
\begin{dfn}\cite{Lod3}
Let $(\mathfrak{g},[.,.])$ be a right (left) Leibniz algebra. A vector space $M$ is called a $\mathfrak{g}$-module if there exist  the following actions of $\mathfrak{g}$ on $M$:
\begin{align*}
[.,.]_{L}:\mathfrak{g}\times M\longrightarrow M\quad ,\quad [.,.]_{R}:M\times \mathfrak{g}\longrightarrow M,
\end{align*}
such that, for the right Leibniz algebra we have the following identities:
\begin{align}
\begin{split}
& 1)\,[[X,Y],m]_{L}=[[X,m]_{L},Y]_{R}+[X,[Y,m]_{L}]_{L},\\	
& 2)\,[[m,Y]_{R},X]_{R}=[[m,X]_{R},Y]_{R}+[m,[Y,X]]_{R},\\
& 3)\,[[Y,m]_{L},X]_{R}=[[Y,X],m]_{L}+[Y,[m,X]_{R}]_{L},\quad\forall X,Y\in \mathfrak{g},\quad\forall m\in M,\label{rightmodule}
\end{split}
\end{align}
while, for the left one we have
\begin{align}
\begin{split}
& 1)\,[m,[X,Y]]_{R}=[[m,X]_{R},Y]_{R}+[X,[m,Y]_{R}]_{L},\\
& 2)\,[X,[m,Y]_{R}]_{L}=[[X,m]_{L},Y]_{R}+[m,[X,Y]]_{R},\\
& 3)\,[X,[Y,m]_{L}]_{L}=[[X,Y],m]_{L}+[Y,[X,m]_{L}]_{L},\quad\forall X,Y\in \mathfrak{g},\quad\forall m\in M.\label{leftmodule}
\end{split}
\end{align}
\end{dfn}
\begin{dfn}\cite{Lod3}
The right (left) Leibniz cohomology $HL^{n}(\mathfrak{g},M)$ of a right (left) Leibniz algebra $\mathfrak{g}$ with a representation $M$ (i.e. $M$ is a $\mathfrak{g}$-module.)  is defined for the complex\linebreak $CL^{n}(\mathfrak{g},M)=Hom(\mathfrak{g}^{\otimes n},M)$ with the Leibniz coboundary map 
$\delta^{n}:CL^{n}(\mathfrak{g},M)\longrightarrow CL^{n+1}(\mathfrak{g},M),$ such that, $\forall\omega\in CL^{n}(\mathfrak{g},M)$ and $X_{1}, . . ., X_{n+1}\in \mathfrak{g}$, we have
\begin{align*}
(\delta^{n}\omega)(X_{1},X_{2}, . . .,  ,X_{n+1})&=[X_{1},\omega(X_{2}, . . .,  X_{n+1})]_{L}\\
&\,+\sum_{i=2}^{n+1}{(-1)^{i}[\omega(X_{1}, . . ., \widehat{X_{i}}, . . ., X_{n+1}),X_{i}]_{R}}\\
&\,+\sum_{1\leq i<j\leq n+1}{(-1)^{j+1}\omega(X_{1}, . . .,  X_{i-1},[X_{i},X_{j}],X_{i+1}, . . ., \widehat{X_{j}}, . . .,  X_{n+1})},
\end{align*}
for the right Leibniz algebra, and
\begin{align*}
(\delta^{n}\omega)(X_{1},X_{2}, . . ., X_{n+1})&=\sum_{i=1}^{n}{[X_{i},\omega(X_{1}, . . ., \widehat{X_{i}}, . . ., X_{n+1})]_{L}}\\
&\,+(-1)^{n+1}[\omega(X_{1}, . . ., X_{n}),X_{n+1}]_{R}\\
&\,+\sum_{1\leq i<j\leq n+1}{(-1)^{i}\omega(X_{1}, . . ., \widehat{ X_{i}}, . . ., X_{j-1},[X_{i},X_{j}], . . ., X_{n+1})},
\end{align*}
for the left one.
\end{dfn}
As an example for the right Leibniz algebra, 
$\forall m\in CL^{0}(\mathfrak{g},M)=C^{0}(\mathfrak{g},M)=M$, we have
\begin{align*}
(\delta^{0}m)(X)=[X,m]_{L},
\end{align*}
and $\forall\omega\in CL^{1}(\mathfrak{g},M)=C^{1}(\mathfrak{g},M)$, we have
\begin{align*}
(\delta^{1}\omega)(X,Y)=[X,\omega(Y)]_{L}+[\omega(X),Y]_{R}-\omega([X,Y]),
\end{align*}
in the same way for the left Leibniz algebra we have the following relations:
\begin{align*}
&(\delta^{0}m)(X)=-[m,X]_{R},\\
&(\delta^{1}\omega)(X,Y)=[X,\omega(Y)]_{L}+[\omega(X),Y]_{R}-\omega([X,Y]).
\end{align*}
\begin{dfn}\cite{Lod3}
Let $\mathfrak{g}$ be a right or left Leibniz algebra, then, 
$\omega\in CL^{1}(\mathfrak{g},M)=C^{1}(\mathfrak{g},M)$ is called $1$-cocycle if 
$\forall X,Y\in\mathfrak{g}$, we have
\begin{align*}
[X,\omega(Y)]_{L}+[\omega(X),Y]_{R}-\omega([X,Y])=0.
\end{align*}
\end{dfn}
\section{Leibniz bialgebras}
\hspace{13.5pt}
Before defining the Leibniz bialgebra let us define especial actions of $\mathfrak{g}$ on $\mathfrak{g}\otimes \mathfrak{g}$. 
Let $\mathfrak{g}$ be a finite dimensional  left Leibniz algebra and $\gamma:\mathfrak{g}\longrightarrow \mathfrak{g}\otimes \mathfrak{g}$ be a linear map.
Also, we denote  the transpose of $\gamma$ by 
$\gamma ^{t}:\mathfrak{g}^{\ast}\otimes \mathfrak{g}^{\ast}\longrightarrow \mathfrak{g}^{\ast}$,
where, $\mathfrak{g}^{\ast}$ is  the dual space of $\mathfrak{g}$. We define the following  actions of $\mathfrak{g}$ on $\mathfrak{g}\otimes \mathfrak{g}$  such that  $\mathfrak{g}\otimes \mathfrak{g}$ be a $\mathfrak{g}$-module:
\begin{itemize}
\item
\begin{align}
\begin{split}
&[,]_{L}:\mathfrak{g}\times(\mathfrak{g}\otimes \mathfrak{g})\longrightarrow (\mathfrak{g}\otimes \mathfrak{g}),\quad [X,Y\otimes Z]_{L}:=(\mathrm{ad}_{X}^{(l)}\otimes 1)(Y\otimes Z),\\
&[,]_{R}:(\mathfrak{g}\otimes \mathfrak{g})\times\mathfrak{g} \longrightarrow (\mathfrak{g}\otimes \mathfrak{g}),\quad [Y\otimes Z,X]_{R}:=(\mathrm{ad}_{X}^{(r)}\otimes 1)(Y\otimes Z).
\label{lorr-r}
\end{split}\hspace{2cm}
\end{align}
\item
\begin{align}
\begin{split}
&[,]_{L}:\mathfrak{g}\times(\mathfrak{g}\otimes \mathfrak{g})\longrightarrow (\mathfrak{g}\otimes \mathfrak{g}),\quad [X,Y\otimes Z]_{L}:=(1\otimes \mathrm{ad}_{X}^{(l)}+ \mathrm{ad}_{X}^{(l)}\otimes 1)(Y\otimes Z),\\
&[,]_{R}:(\mathfrak{g}\otimes \mathfrak{g})\times\mathfrak{g} \longrightarrow (\mathfrak{g}\otimes \mathfrak{g}),\quad [Y\otimes Z,X]_{R}:=0.\label{l-lorr}
\end{split}
\end{align}
\item
\begin{align}
\begin{split}
&[,]_{L}:\mathfrak{g}\times(\mathfrak{g}\otimes \mathfrak{g})\longrightarrow (\mathfrak{g}\otimes \mathfrak{g}),\quad [X,Y\otimes Z]_{L}:=(1\otimes \mathrm{ad}_{X}^{(l)})(Y\otimes Z),\\
&[,]_{R}:(\mathfrak{g}\otimes \mathfrak{g})\times\mathfrak{g} \longrightarrow (\mathfrak{g}\otimes \mathfrak{g}),\quad [Y\otimes Z,X]_{R}:=(1\otimes \mathrm{ad}_{X}^{(r)})(Y\otimes Z),
\label{lorr-l}
\end{split}\hspace{2cm}
\end{align}
\end{itemize}
where, in the above relations "$1$" is the identity map on $\mathfrak{g}$. Note that if
$\mathfrak{g}$ be a finite dimensional  right Leibniz algebra one can use the actions
\eqref{lorr-r}, \eqref{lorr-l} and the following action:
\begin{align}
\begin{split}
&[,]_{L}:\mathfrak{g}\times(\mathfrak{g}\otimes \mathfrak{g})\longrightarrow (\mathfrak{g}\otimes \mathfrak{g}),\quad [X,Y\otimes Z]_{L}:=0,\label{r-lorr}\\
&[,]_{R}:(\mathfrak{g}\otimes \mathfrak{g})\times\mathfrak{g} \longrightarrow (\mathfrak{g}\otimes \mathfrak{g}),\quad [Y\otimes Z,X]_{R}:=(1\otimes \mathrm{ad}_{X}^{(r)}+ \mathrm{ad}_{X}^{(r)}\otimes 1)(Y\otimes Z).
\end{split}
\end{align}

Now, using these actions we define the Leibniz bialgebra, as follows.
\begin{dfn}
 A {\textit{Leibniz bialgebra}} $(\mathfrak{g},\gamma)$ is a Leibniz algebra $\mathfrak{g}$ with a linear map (cocommutator) $\gamma:\mathfrak{g}\longrightarrow \mathfrak{g}\otimes \mathfrak{g}$, such that
\begin{itemize}
\item
$\gamma$ is a $1$-cocycle on $\mathfrak{g}$ with values in $\mathfrak{g}\otimes \mathfrak{g}$ \footnote{Note that $\mathfrak{g}$ acts on $\mathfrak{g}\otimes \mathfrak{g}$ from left and right such that $\mathfrak{g}\otimes \mathfrak{g}$ becomes a $\mathfrak{g}$-module.}.
\begin{align}
[X,\gamma(Y)]_{L}+[\gamma(X),Y]_{R}-\gamma([X,Y])=0,\label{1-cocycle}
\end{align}
\item
$\gamma ^{t}:\mathfrak{g}^{\ast}\otimes \mathfrak{g}^{\ast}\longrightarrow\mathfrak{g}^{\ast}$
 defines a Leibniz bracket on $\mathfrak{g}^{\ast}$.
\end{itemize}
\end{dfn}
With the notation 
$[\xi,\eta]_{\ast}:=\gamma^{t}(\xi\otimes\eta)$ for any $\xi,\eta\in\mathfrak{g}^{\ast}$ and $X\in\mathfrak{g}$ we have
\begin{align}
\langle [\xi,\eta]_{\ast},X\rangle=\langle\gamma^{t}(\xi\otimes \eta),X\rangle=\langle\gamma(X),\xi\otimes \eta\rangle, \label{pairing}
\end{align}
where, $\langle ., .\rangle$ is the natural pairing between the spaces $\mathfrak{g}$ and $\mathfrak{g}^{\ast}$. For the left Leibniz algebra $\mathfrak{g}$ with the actions \eqref{lorr-r}-\eqref{lorr-l} on $\mathfrak{g}\otimes \mathfrak{g}$; the $1$-cocycle condition \eqref{1-cocycle} can be rewritten in the following forms:
\begin{align}
& \gamma_{(lr-r)}[X,Y]:=(\mathrm{ad}_{X}^{(l)}\otimes 1)\gamma_{(lr-r)}(Y)+(\mathrm{ad}_{Y}^{(r)}\otimes 1)\gamma_{(lr-r)}(X),\label{1-cocyclelorr-r}\\
& \gamma_{(l-lr)}[X,Y]:=(1\otimes \mathrm{ad}_{X}^{(l)}+ \mathrm{ad}_{X}^{(l)}\otimes 1)\gamma_{(l-lr)}(Y),\label{1-cocyclel-lorr}\\
&\gamma_{(lr-l)}[X,Y]:=(1\otimes \mathrm{ad}_{X}^{(l)})\gamma_{(lr-l)}(Y)+(1\otimes \mathrm{ad}_{Y}^{(r)})\gamma_{(lr-l)}(X).\label{1-cocyclelorr-l}
\end{align}
Note that for the right Leibniz algebra the $1$-cocycle condition \eqref{1-cocycle} can be rewitten as \eqref{1-cocyclelorr-r}, \eqref{1-cocyclelorr-l} and also as the following
\begin{align}
\gamma_{(r-lr)}[X,Y]:=(1\otimes \mathrm{ad}_{Y}^{(r)}+\mathrm{ad}_{Y}^{(r)}\otimes 1)\gamma_{(r-lr)}(X).\label{1-cocycler-lorr}
\end{align}
In the above definition, if $\mathfrak{g}$ is a left Leibniz algebra then $\mathfrak{g}^*$ can be a left or right Leibniz algebra and the same is true for the right Leibniz algebra $\mathfrak{g}$. The notation $\gamma_{(lr-r)}$ shows that $\mathfrak{g}$ can be left or right Leibniz algebra and $\mathfrak{g^*}$ is a right Leibniz algebra or the notation
$\gamma_{(l-lr)}$ state that  $\mathfrak{g}$ is a left Leibniz algebra and $\mathfrak{g^*}$ can be left or right Leibniz algebra and so on.
\begin{rem}
If the Leibniz algebra $\mathfrak{g}$ be a Lie algebra, we have $\mathrm{ad}^{(l)}=-\mathrm{ad}^{(r)}=\mathrm{ad}$ such that, any cases of \eqref{1-cocyclelorr-r}-\eqref{1-cocyclelorr-l} do not satisfy  in $1$-cocycle condition of Lie case, directly. In this case we must consider the composition of \eqref{1-cocyclelorr-r}-\eqref{1-cocyclelorr-l}, such that, $\mathfrak{g}$ and $\mathfrak{g}^{\ast}$ be a Lie bialgebra, for example, the composition of \eqref{1-cocyclelorr-r} and \eqref{1-cocyclelorr-l}
  such that, for Lie algebra case the $1$-cocycle condition  can be written as follows:
\begin{align*}
\gamma[X,Y]=[X,\gamma(Y)]-[Y,\gamma(X)]=(1\otimes \mathrm{ad}_{X}+\mathrm{ad}_{X}\otimes 1)\gamma(Y)-(1\otimes \mathrm{ad}_{Y}+\mathrm{ad}_{Y}\otimes 1)\gamma(X),
\end{align*}
where, $\mathrm{ad}^{(2)}:=1\otimes \mathrm{ad}+\mathrm{ad}\otimes 1$ defines the left $\mathfrak{g}$-module structure on $\mathfrak{g}\otimes \mathfrak{g}$. Also, one can consider
\begin{align*}
\gamma[X,Y]=[\gamma(X),Y]-[\gamma(Y),X],
\end{align*}
such that, $[.,.]$ defines the right $\mathfrak{g}$-module structure on $\mathfrak{g}\otimes \mathfrak{g}$.
\end{rem}
We present all the propositions and examples for the left Leibniz algebras, and all of them for the right Leibniz algebras are quite similar and proofs are analogous.
\begin{prop}
If $(\mathfrak{g},\gamma)$ be a  Leibniz bialgebra, and $\mu$ be the left Leibniz bracket on $\mathfrak{g}$, then
$(\mathfrak{g}^{\ast},\mu^{t})$ will be a Leibniz bialgebra, where $\gamma^{t}$ is the (left or right)Leibniz bracket of $\mathfrak{g}^{\ast}$.
\end{prop}
\begin{proof}
According to the which one of the $1$-cocycle conditions hold; the Leibniz algebra $\mathfrak{g}^{\ast}$ can be right or left.
We investigate this proposition, as follows:\newline
Suppose \eqref{1-cocyclelorr-r} holds, then from \eqref{pairing} we have
\begin{align}
\begin{split}
&\langle [\xi,\eta]_{\ast},[X,Y]\rangle=\langle \xi\otimes \eta,\gamma_{(lr-r)}[X,Y]\rangle
\\
&=\langle \xi\otimes \eta,(\mathrm{ad}_{X}^{(l)}\otimes 1)\gamma_{(lr-r)}(Y)\rangle +\langle \xi\otimes \eta,( \mathrm{ad}_{Y}^{(r)}\otimes 1)\gamma_{(lr-r)}(X)\rangle.\label{1'}
\end{split}
\end{align}
We now need to define the right and left coadjoint action of a Leibniz algebra $\mathfrak{g}$ on the dual vector space $\mathfrak{g}^{\ast}$. Let $\mathfrak{g}$ be a left Leibniz algebra and $\mathfrak{g}^{\ast}$ be its dual vector space, then for $X\in\mathfrak{g}$ in \cite{Uch}, Uchino defined two actions of $\mathfrak{g}$ on $\mathfrak{g}^{\ast}$ 
\begin{align*}
&{\mathrm{ad}_{X}^{\ast}}^{(l)}:\mathfrak{g}^{\ast}\longrightarrow \mathfrak{g}^{\ast}\qquad,\qquad
\langle {\mathrm{ad}_{X}^{\ast}}^{(l)}\xi,Y\rangle:=-\langle\xi,\mathrm{ad}_{X}^{(l)}Y\rangle,\cr
&{\mathrm{ad}_{X}^{\ast}}^{(r)}:\mathfrak{g}^{\ast}\longrightarrow \mathfrak{g}^{\ast}\qquad,\qquad
\langle {\mathrm{ad}_{X}^{\ast}}^{(r)}\xi,Y\rangle:=\langle\xi,\mathrm{ad}_{X}^{(l)}Y+\mathrm{ad}_{X}^{(r)}Y\rangle,
\end{align*}
for any $Y\in\mathfrak{g}$ and $\xi\in \mathfrak{g}^\ast$ such that $\mathfrak{g}^{\ast}$ be a $\mathfrak{g}$-module.
 Using these relations, \eqref{1'} can be rewritten as
\begin{align}
&\langle [\xi,\eta]_{\ast},[X,Y]\rangle+\langle [{\mathrm{ad}_{X}^{\ast}}^{(l)}\xi,\eta]_{\ast},Y\rangle-\langle[{\mathrm{ad}_{Y}^{\ast}}^{(r)}\xi,\eta]_{\ast},X\rangle-\langle[{\mathrm{ad}_{Y}^{\ast}}^{(l)}\xi,\eta]_{\ast},X\rangle=0.
\label{1''}
\end{align}
Similarly, $\mathfrak{g}^{\ast}$ can act on $\mathfrak{g}$ from left and right. Let $\mathfrak{g}^{\ast}$ be a left Leibniz algebra, then for any
$\xi,\eta\in\mathfrak{g}^{\ast}$ and $X\in\mathfrak{g}$ we have:
\begin{align*}
&{\mathrm{ad}_{\xi}^{\ast}}^{(l)}:\mathfrak{g}\longrightarrow \mathfrak{g}\cong {\mathfrak{g}^{\ast}}^{\ast}\quad,\quad
({\mathrm{ad}_{\xi}^{\ast}}^{(l)}X)(\eta)=\langle {\mathrm{ad}_{\xi}^{\ast}}^{(l)}X,\eta\rangle:=-\langle X,\mathrm{ad}_{\xi}^{(l)}\eta\rangle,\\
&{\mathrm{ad}_{\xi}^{\ast}}^{(r)}:\mathfrak{g}\longrightarrow \mathfrak{g}\cong {\mathfrak{g}^{\ast}}^{\ast}\quad,\quad
({\mathrm{ad}_{\xi}^{\ast}}^{(r)}X)(\eta)=\langle {\mathrm{ad}_{\xi}^{\ast}}^{(r)}X,\eta\rangle:=\langle X,\mathrm{ad}_{\xi}^{(l)}\eta+\mathrm{ad}_{\xi}^{(r)}\eta\rangle.
\end{align*}
If $\mathfrak{g}^{\ast}$ be a right Leibniz algebra, then for every
$\xi,\eta\in\mathfrak{g}^{\ast}$ and $X\in\mathfrak{g}$ we have:
\begin{align*}
&{\mathrm{ad}_{\xi}^{\ast}}^{(l)}:\mathfrak{g}\longrightarrow \mathfrak{g}\cong {\mathfrak{g}^{\ast}}^{\ast}\quad,\quad
({\mathrm{ad}_{\xi}^{\ast}}^{(l)}X)(\eta)=\langle {\mathrm{ad}_{\xi}^{\ast}}^{(l)}X,\eta\rangle:=\langle X,\mathrm{ad}_{\xi}^{(l)}\eta+\mathrm{ad}_{\xi}^{(r)}\eta\rangle,\\
&{\mathrm{ad}_{\xi}^{\ast}}^{(r)}:\mathfrak{g}\longrightarrow \mathfrak{g}\cong {\mathfrak{g}^{\ast}}^{\ast}\quad,\quad
({\mathrm{ad}_{\xi}^{\ast}}^{(r)}X)(\eta)=\langle {\mathrm{ad}_{\xi}^{\ast}}^{(r)}X,\eta\rangle:=-\langle X,\mathrm{ad}_{\xi}^{(r)}\eta\rangle.
\end{align*}
Using the above relations, \eqref{1''} 
can be rewritten as
\begin{align}
&\langle [\xi,\eta]_{\ast},[X,Y]\rangle-\langle {\mathrm{ad}_{X}^{\ast}}^{(l)}\xi,{\mathrm{ad}_{\eta}^{\ast}}^{(r)}Y\rangle
+\langle {\mathrm{ad}_{Y}^{\ast}}^{(r)}\xi,{\mathrm{ad}_{\eta}^{\ast}}^{(r)}X\rangle+\langle {\mathrm{ad}_{Y}^{\ast}}^{(l)}\xi,{\mathrm{ad}_{\eta}^{\ast}}^{(r)}X\rangle=0,\label{1'''}
\end{align}
	or
\begin{align*}
&\langle [\xi,\eta]_{\ast},\mu(X\otimes Y)\rangle+\langle\xi,[X,{\mathrm{ad}_{\eta}^{\ast}}^{(r)}Y]\rangle
+\langle\xi,[{\mathrm{ad}_{\eta}^{\ast}}^{(r)}X,Y]\rangle=0,
\end{align*}
where $\mu$ is the Leibniz bracket on $\mathfrak{g}$ and $\mu^{t}$ is the cocommutator on $\mathfrak{g}^{\ast}$ i.e. $\mu^{t}:\mathfrak{g}^{\ast}\longrightarrow \mathfrak{g}^{\ast}\otimes \mathfrak{g}^{\ast}$. Therefore, we have
\begin{align*}
&\langle\mu^{t}[\xi,\eta]_{\ast},X\otimes Y\rangle-\langle(\mathrm{ad}_{\eta}^{(r)}\otimes 1+1\otimes \mathrm{ad}_{\eta}^{(r)})\mu^{t}(\xi),X\otimes Y\rangle=0,
\end{align*}
or
\begin{align}
&(\mathrm{ad}_{\eta}^{(r)}\otimes 1+1\otimes \mathrm{ad}_{\eta}^{(r)})\mu^{t}(\xi)=\mu^{t}[\xi,\eta]_{\ast}.\label{dual1-cocycle}
\end{align}
However, this relation is the $1$-cocycle condition \eqref{1-cocycler-lorr} for the Leibniz bialgebra $(\mathfrak{g}^{\ast},\mu^{t}),$ such that, it  indicates that the action of $\mathfrak{g}^{\ast}$ on $\mathfrak{g}^{\ast}\otimes \mathfrak{g}^{\ast}$ like the case \eqref{r-lorr}, i.e., {\textit{$\mathfrak{g}^{\ast}$ is a right Leibniz algebra.}}\newline

In the same way,
for the left Leibniz algebra $(\mathfrak{g},\mu)$ if one uses 
 $\gamma_{(l-lr)}[X,Y]$, then
\begin{align*}
(1\otimes \mathrm{ad}_{\xi}^{(l)})\mu^{t}(\eta)+ (1\otimes \mathrm{ad}_{\eta}^{(r)})\mu^{t}(\xi)=\mu^{t}[\xi,\eta]_{\ast},
\end{align*}
instead of \eqref{dual1-cocycle}. This shows that, $\mu^{t}$ is a $1$-cocycle condition \eqref{1-cocyclelorr-l} for the Leibniz bialgebra $(\mathfrak{g}^{\ast},\mu^{t})$ indicating the action of $\mathfrak{g}^{\ast}$ on
$\mathfrak{g}^{\ast}\otimes\mathfrak{g}^{\ast}$ as the case \eqref{lorr-l}, i.e., $\mathfrak{g}^{\ast}\otimes \mathfrak{g}^{\ast}$ is a $\mathfrak{g}^{\ast}$-module and {\textit{$\mathfrak{g}^{\ast}$ is a left or a right Leibniz algebra.}}\newline
Finally, for a left
Leibniz algebra $(\mathfrak{g},\mu)$ with  $\gamma_{(lr-l)}[X,Y]$
, we have
\begin{align*}
(1\otimes \mathrm{ad}_{\xi}^{(l)}+\mathrm{ad}_{\xi}^{(l)}\otimes 1)\mu^{t}(\eta)=\mu^{t}[\xi,\eta]_{\ast},
\end{align*}
instead of \eqref{dual1-cocycle}, showing that $\mu^{t}$ is a $1$-cocycle condition \eqref{1-cocyclel-lorr} for the Leibniz bialgebra $(\mathfrak{g}^{\ast},\mu^{t}),$ such that, it shows that the action of $\mathfrak{g}^{\ast}$ on $\mathfrak{g}^{\ast}\otimes \mathfrak{g}^{\ast}$ as the case \eqref{l-lorr}; i.e. $\mathfrak{g}^{\ast}\otimes \mathfrak{g}^{\ast}$ is a $\mathfrak{g}^{\ast}$-module and {\textit{$\mathfrak{g}^{\ast}$ is a left Leibniz algebra.}}
	\end{proof}\newline	
A Leibniz bialgebra $(\mathfrak{g},\gamma)$ can also be denoted by $(\mathfrak{g},\mathfrak{g}^{\ast})$. If we choose $(\{X_i\},f_{ij}\,^{k})$ and $(\{\widetilde{X}^{i}\},\widetilde{f}^{ij}\,_{k})$  as the basis and structure constants of  Leibniz algebra $\mathfrak{g}$ and $\mathfrak{g}^{\ast}$, respectively, it follows that\footnote{Here we use the Einstein summation convention, i.e., we have the summation  over  the upper and lower indices.}
\begin{align}
[X_{i},X_{j}]=f_{ij}\,^{k}X_{k},\quad [\widetilde{X}^{i},\widetilde{X}^{j}]=\widetilde{f}^{ij}\,_{k}\widetilde{X}^{k},\label{constantg&dual}
\end{align}
then, from \eqref{constantg&dual} and \eqref{pairing}, we have
\begin{align}
\gamma (X_{k})=\widetilde{f}^{ij}\,_{k} X_{i}\otimes X_{j}.\label{cocommutator}
\end{align}
At the end of this section we should mention that there is no Manin triple related to the Leibniz bialgebra (as in Lie bialgebra case), because the $1$-cocycle conditions in $\mathfrak{g}$ and $\mathfrak{g}^\ast$ are different and unsymmetrical.
\section{Coboundary Leibniz bialgebras, classical Yang-Baxter equation and $r$-matrices}
In this section, we will define coboundary Leibniz bialgebra in a way that the Leibniz bialgebra structures on a Leibniz algebra $\mathfrak{g}$ is defined by a cocycle $\delta^{0}r$ which is the coboundary of an element
$r\in\mathfrak{g}\otimes \mathfrak{g}$ and $r$  is called $r$-matrices\footnote{In the following we use and generalize the similar method for Lie bialgebra as in \cite{Sch} for the Leibniz bialgebra case.}. Also, we define the classical Yang-Baxter type equation for the Leibniz bialgebra and show that it is a sufficient condition for $\delta^{0}r$ to define a Leibniz bracket on $\mathfrak{g}^{\ast}$. Notice that in this case we suppose $\mathfrak{g}$ and $\mathfrak{g}^{\ast}$ to be both right or left Leibniz algebras.
\begin{dfn}
A Leibniz bialgebra $(\mathfrak{g},\gamma)$ is called {\textit{ coboundary Leibniz bialgebra}} if the cocommutator $\gamma$ be a $1$-coboundary, i.e., if there exists an element $r\in\mathfrak{g}\otimes\mathfrak{g}$, such that (if $\mathfrak{g}$ be a left Leibniz algebra)
\begin{itemize}
\item
for $1$-cocycle \eqref{1-cocyclelorr-r}
\begin{align}
\gamma_{(lr-r)}(X)=-(\mathrm{ad}_{X}^{(r)}\otimes 1)(r),\quad\forall X\in\mathfrak{g}\label{cocommutator1'-left}
\end{align}
\item 
for $1$-cocycle 
and \eqref{1-cocyclel-lorr}
\begin{align}
\gamma_{(l-lr)}(X)=0,\quad\forall X\in\mathfrak{g}
\end{align}
\item
for $1$-cocycle \eqref{1-cocyclelorr-l}
\begin{align}
\gamma_{(lr-l)}(X)=-(1\otimes \mathrm{ad}_{X}^{(r)})(r),\quad\forall X\in\mathfrak{g}\label{cocommutator4'-left}
\end{align}
\end{itemize}
\end{dfn}
In terms of the basis $\{X_{i}\}$ for the Leibniz algebra $\mathfrak{g}$, we have, $r=r^{ij} X_{i}\otimes X_{j}$. Then, using \eqref{constantg&dual} and \eqref{cocommutator}, the cocommutators 
 \eqref{cocommutator1'-left}, 
and \eqref{cocommutator4'-left} can be rewritten, respectively, as follows
\begin{align}
&\widetilde{f}^{kj}\,_{m}=-r^{ij}f_{im}\,^{k}\label{rewritecocomm1'-left},\\
&\widetilde{f}^{ik}\,_{m}=-r^{ij}f_{jm}\,^{k}.\label{rewritecocomm4'-left}
\end{align}
For any element $r\in\mathfrak{g}\otimes \mathfrak{g}$ we associate the map
$\underline{r}:\mathfrak{g}^{\ast}\longrightarrow \mathfrak{g}$ defined as
\begin{align}
\underline{r}(\xi)(\eta)=r(\xi,\eta)=\langle \eta,\underline{r}\xi\rangle=\langle \underline{r}^{t}\eta,\xi\rangle,\qquad\forall \xi,\eta\in\mathfrak{g}^{\ast},\label{rbar}
\end{align}
 which $\underline{r}^{t}:\mathfrak{g}^{\ast}\longrightarrow \mathfrak{g}$ denotes the transpose of the map $\underline{r}$. \newline
Now, we prove the following propositions when $\mathfrak{g}$ and $\mathfrak{g}^{\ast}$ are both left Leibniz algebras. When $\mathfrak{g}$ and $\mathfrak{g}^{\ast}$ are both right Leibniz algebras, the results  will be given in  the corollary \ref{cor4.5}. The proof of this case is similar.
\begin{prop}
	If $(\mathfrak{g},\gamma)$ be a Leibniz bialgebra such that $\mathfrak{g}$ and $\mathfrak{g}^{\ast}$ be both left Leibniz algebra and \eqref{cocommutator1'-left}
	is valid; then $\forall\xi,\eta\in\mathfrak{g}^{\ast}$ we will have\footnote{Note that $[\xi,\eta]_{\ast}=\gamma^{t}(\xi,\eta)$, i.e., when $\gamma=\delta^{0}r$, we use $[\xi,\eta]^{r}$ instead of $[\xi,\eta]_{\ast}$.}
	\begin{align}
	\gamma^{t}(\xi\otimes\eta)=[\xi,\eta]_{\ast}=[\xi,\eta]^{r}={\mathrm{ad}^{\ast}}^{(l)}_{\underline{r}(\xi)}\eta.\label{r-bracketforleft}
	\end{align}
\end{prop}
\begin{proof}	
From \eqref{cocommutator1'-left} and \eqref{pairing} for any $X\in \mathfrak{g}$ we have
\begin{align*}
\langle \gamma^{t}(\xi\otimes\eta),X\rangle&=\langle (\delta^{0}r)^{t}(\xi\otimes\eta),X\rangle
=\langle \xi\otimes\eta,(\delta^{0}r)(X)\rangle=-\langle \xi\otimes \eta,(1\otimes\mathrm{ad}_{X}^{(r)})(r)\rangle\\
&=-\langle\xi\otimes{\mathrm{ad}^{\ast}}^{(r)}_{X}\eta,r\rangle=-r(\xi\otimes {\mathrm{ad}^{\ast}}^{(r)}_{X}\eta)=-\underline{r}(\xi)({\mathrm{ad}^{\ast}}^{(r)}_{X}\eta)\\
& =-\langle \underline{r}\xi,{\mathrm{ad}^{\ast}}^{(r)}_{X}\eta,r\rangle=-
\langle {\mathrm{ad}^{\ast}}^{(r)}_{X}\underline{r}\xi,\eta\rangle\\
&=-\langle\mathrm{ad}_{\underline{r}\xi}^{(l)}X,\eta\rangle=
\langle X,{\mathrm{ad}^{\ast}}^{(l)}_{\underline{r}\xi}\eta\rangle
\end{align*}
so that,	
\begin{align*}
\gamma^{t}(\xi\otimes\eta)=[\xi,\eta]^{r}={\mathrm{ad}^{\ast}}^{(l)}_{\underline{r}(\xi)}\eta.
	\end{align*}
\end{proof}
\begin{dfn}
Let $(\mathfrak{g},\gamma)$ be a coboudary Leibniz bialgebra such that $\mathfrak{g}$	and
$\mathfrak{g}^{\ast}$ be both left Leibniz algebras. We introduce the algebraic
Schouten bracket of an element $r\in \mathfrak{g}\otimes\mathfrak{g}$ with itself, denoted by
$[[r,r]]\in\otimes^3{g}$ as follows:
\begin{align}
[[r,r]](\xi,\eta,\zeta)=\langle\eta,[\underline{r}\xi,\underline{r}^{t}\zeta]\rangle+\langle\zeta,[\underline{r}\xi,\underline{r}\eta]\rangle.\label{left-yang-baxter-asli}
\end{align}	
\end{dfn}	
\begin{prop}	
Let	$(\mathfrak{g},\gamma)$ be a coboundary Leibniz bialgebra such that $\mathfrak{g}$	and $\mathfrak{g}^{\ast}$ are both left Leibniz algebras. A necessary and sufficient condition for $\gamma=\delta^{0}r$ (with $r\in\mathfrak{g}\otimes\mathfrak{g}$) to define a left Leibniz bracket on $\mathfrak{g}^{\ast}$ is that the Schouten bracket $[[r,r]]$  be $\mathrm{ad}$-invariant.	
\end{prop}
\begin{proof}
For any $X\in\mathfrak{g}$ and $\forall\xi,\eta,\zeta\in\mathfrak{g}^{\ast}$ the left Leibniz identity for $\mathfrak{g}^{\ast}$ by use of \eqref{r-bracketforleft} can be written as follows:
\begin{align*}
&\langle [\xi,[\eta,\zeta]^{r}]^{r}-[[\xi,\eta]^{r},\zeta]^{r}-[\eta,[\xi,\zeta]^{r}]^{r},X\rangle\\
&=\langle [\xi,{\mathrm{ad}^{\ast}}^{(l)}_{\underline{r}^{t}(\eta)}\zeta]^{r}-
[{\mathrm{ad}^{\ast}}^{(l)}_{\underline{r}(\xi)}\eta,\zeta]^{r}-
[\eta,{\mathrm{ad}^{\ast}}^{(l)}_{\underline{r}(\xi)}\zeta]^{r},X\rangle\\
&=\langle {\mathrm{ad}^{\ast}}^{(l)}_{\underline{r}(\xi)} {\mathrm{ad}^{\ast}}^{(l)}_{\underline{r}(\eta)}\zeta-{\mathrm{ad}^{\ast}}^{(l)}_{\underline{r}{\mathrm{ad}^{\ast}}^{(l)}_{\underline{r}(\xi)}\eta}(\zeta)-{\mathrm{ad}^{\ast}}^{(l)}_{\underline{r}(\eta)}{\mathrm{ad}^{\ast}}^{(l)}_{\underline{r}(\xi)}\zeta,X\rangle\\
&=\langle {\mathrm{ad}^{\ast}}^{(l)}_{[\underline{r}(\xi),\underline{r}(\eta)]}\zeta,X \rangle-\langle{\mathrm{ad}^{\ast}}^{(l)}_{\underline{r}({\mathrm{ad}^{\ast}}^{(l)}_{\underline{r}(\xi)}\eta)}\zeta,X\rangle\\
&=-\langle\zeta,[[\underline{r}(\xi),\underline{r}(\eta)],X]\rangle+\langle\zeta,({\mathrm{ad}}^{(r)}_X\underline{r}({\mathrm{ad}^{\ast}}^{(l)}_{\underline{r}(\xi)}\eta)\rangle\\
&+\langle\zeta,\mathrm{ad}^{(l)}_X\underline{r}({\mathrm{ad}^{\ast}}^{(l)}_{\underline{r}(\xi)}\eta)\rangle-\langle\zeta,\mathrm{ad}^{(l)}_X\underline{r}({\mathrm{ad}^{\ast}}^{(l)}_{\underline{r}(\xi)}\eta)\rangle\\
&=-\delta^{0}([[r,r]])(X)(\xi,\eta,\zeta)=(1\otimes 1\otimes \mathrm{ad}^{(r)}_{X})[[r,r]]
(\xi,\eta,\zeta),
\end{align*}
obviously, a sufficient condition for $\gamma=\delta^{0}r$ to define a Leibniz bracket $[\xi,\eta]^{r}=(\delta^{0}r)^{t}(\xi\otimes\eta)$ is when $[[r,r]]$ is $ad$-invariant, i.e., $(1\otimes 1\mathrm{ad}_{X}^{(r)})[[r,r]]=0$.
\end{proof}	
\begin{cor}\label{cor4.5}
If $(\mathfrak{g},\gamma)$ be a coboudary Leibniz bialgebra such that $\mathfrak{g}$ and $\mathfrak{g}^{\ast}$ be both right Leibniz algebra and $\gamma=\delta^{0}r$ with $r\in\mathfrak{g}\otimes\mathfrak{g}$, then we will have
\begin{align}
&[\xi,\eta]^{r}=-{\mathrm{ad}^{\ast}}^{(r)}_{\underline{r}^{t}(\eta)}\xi,\label{r-bracketforright}\\
&[[r,r]](\xi,\eta,\zeta)=\langle \xi,[\underline{r}^{t}\eta,\underline{r}^{t}\zeta]\rangle+\langle\eta,[\underline{r}\xi,\underline{r}^{t}\zeta]\rangle,\quad\forall\xi,\eta,\zeta\in\mathfrak{g}^{\ast}.\label{right-yang-baxter-asli}
\end{align}
\end{cor}
\begin{dfn}
As for the Lie algebra case we call the condition $[[r,r]]=0$ with $[[r,r]]$ being $\mathrm{ad}$-invariant, the  \textbf{classical Yang-Baxter equation} and \textbf{generalized Yang-Baxter equation}, respectively. 	
\end{dfn}
Let $(\mathfrak{g},\gamma)$ be a Leibniz bialgebra such that $\mathfrak{g}$ and $\mathfrak{g}^{\ast}$ are both left Leibniz algebras. Next, suppose that $\{X_i\}$ and $\{\widetilde{X}^{i}\}$ be the basis of the $\mathfrak{g}$ and $\mathfrak{g}^{\ast}$, respectively. Then, we have
\begin{align*}
r(\widetilde{X}^{m}\otimes\widetilde{X}^{n})=r^{ij}(X_i\otimes X_j)(\widetilde{X}^{m}\otimes\widetilde{X}^{n})=r^{ij}\widetilde{X}^{m}(X_i)
\widetilde{X}^{n}(X_j)=r^{mn},
\end{align*}
and from \eqref{rbar} we see that
\begin{align*}
&\underline{r}^{t}\widetilde{X}^{n}=r^{in}X_i,\\
&\underline{r}\widetilde{X}^{m}=r^{mi}X_i.
\end{align*}
In this way one can rewrite \eqref{left-yang-baxter-asli} in terms of the structure constants  as follows:
\begin{align*}
	&[[r,r]](\widetilde{X}^{m},\widetilde{X}^{n},\widetilde{X}^{p})=-\langle \widetilde{X}^{n},[\underline{r}\widetilde{X}^{m},\underline{r}^{t}\widetilde{X}^{p}]\rangle-\langle \widetilde{X}^{p},[\underline{r}\widetilde{X}^{m},\underline{r}\widetilde{X}^{n}]\rangle\\
	&=-\langle \widetilde{X}^{n},[{r}^{mi}{X}_{i},{r}^{jp}{X}_{j}]\rangle-\langle \widetilde{X}^{p},[{r}^{mi}{X}^{i},{r}^{nj}{X}_{j}]\rangle
	=-r^{mi}r^{jp}f_{ij}\,^{n}-r^{mi}r^{nj}f_{ij}\,^{p},
\end{align*}
now by setting
\begin{align*}
	r_{21}:=r^{ij}X_{j}\otimes X_{i}\otimes 1,\quad r_{31}:=r^{ij}X_{j}\otimes 1\otimes X_{i},\quad r_{32}:=r^{ij}1\otimes X_{j}\otimes X_{i},
\end{align*}
in $\otimes^{3} \mathfrak{g}$, one can define
\begin{align*}
	[r_{21},r_{31}]&=[r^{ij}X_{j}\otimes X_{i}\otimes 1, r^{lk} X_{k}\otimes 1\otimes X_{l}]=-r^{ij}r^{lk} \mathrm{ad}^{(r)}_{X_{j}}X_{k}\otimes X_{i}\otimes X_{l}\\
	&= r^{ij}r^{lk} [X_{k},X_{j}]\otimes X_{i}\otimes X_{l}=-r^{ij}r^{lk}f_{kj}\,^{s}X_{s} \otimes X_{i}\otimes X_{l},
\end{align*}
\begin{align*}
	[r_{21},r_{32}]&=[r^{ij}X_{j}\otimes X_{i}\otimes 1,r^{lk}1\otimes X_{k}\otimes X_{l}]= -r^{ij}r^{lk}X_{j}\otimes \mathrm{ad}^{(r)}_{X_{i}}X_{k}\otimes X_{l}\\
	&=-r^{ij}r^{lk}X_{j}\otimes [X_{k},X_{i}]\otimes X_{l}=-r^{ij}r^{lk}f_{ki}\,^{s}X_{j}\otimes X_{s}\otimes X_{l},
\end{align*}
\begin{align*}
	[r_{31},r_{32}]&=[r^{ij}X_{j}\otimes 1\otimes X_{i},r^{lk}1\otimes X_{k}\otimes X_{l}]=-r^{ij}r^{lk}X_{j}\otimes X_{k}\otimes \mathrm{ad}^{(r)}_{X_{i}}X_{l}\\
	&=- r^{ij}r^{lk}X_{j}\otimes X_{k}\otimes [X_{l},X_{i}]=-r^{ij}r^{lk}f_{li}\,^{s}X_{j}\otimes X_{k}\otimes X_{s},
\end{align*}
\normalfont
therefore, we obtain
\begin{align*}
	[r_{21},r_{31}](\widetilde{X}^{m}\otimes \widetilde{X}^{n}\otimes \widetilde{X}^{p})
	=-r^{nj}r^{pk}f_{kj}\,^{m}
\end{align*}
\begin{align*}
	[r_{21},r_{32}](\widetilde{X}^{m}\otimes \widetilde{X}^{n}\otimes \widetilde{X}^{p})
	=-r^{im}r^{pk}f_{ki}\,^{n}
\end{align*}
\begin{align*}
	[r_{31},r_{32}](\widetilde{X}^{m}\otimes \widetilde{X}^{n}\otimes \widetilde{X}^{p})
	=-r^{im}r^{ln}f_{li}\,^{p}
\end{align*}		
and consquently, we have
\begin{align*}
	[[r,r]](\widetilde{X}^{m}\otimes \widetilde{X}^{n}\otimes \widetilde{X}^{p})=[r_{21},r_{31}](\widetilde{X}^{m}\otimes \widetilde{X}^{n}\otimes \widetilde{X}^{p})+[r_{21},r_{32}](\widetilde{X}^{m}\otimes \widetilde{X}^{n}\otimes \widetilde{X}^{p}),
\end{align*}
or
\begin{align*}
	[[r,r]]=[r_{21},r_{31}]+[r_{21},r_{32}].
\end{align*}
Note that, we have proved the following proposition:
\begin{prop}
	If $(\mathfrak{g},\gamma)$ be a Leibniz bialgebra such that $\mathfrak{g}$ and $\mathfrak{g}^{\ast}$ be both left Leibniz algebras, one can rewrite
	\textbf{Yang-Baxter equation} \eqref{left-yang-baxter-asli} in terms of the structure constants  as follows:
	\begin{align*}
		[[r,r]]=[r_{21},r_{31}]+[r_{21},r_{32}].
	\end{align*}
\end{prop}
\begin{proof}
	From \eqref{left-yang-baxter-asli} we have
	\begin{align*}
		&[[r,r]](\widetilde{X}^{m},\widetilde{X}^{n},\widetilde{X}^{p})=-\langle \widetilde{X}^{n},[\underline{r}\widetilde{X}^{m},\underline{r}^{t}\widetilde{X}^{p}]\rangle-\langle \widetilde{X}^{p},[\underline{r}\widetilde{X}^{m},\underline{r}\widetilde{X}^{n}]\rangle\\
		&=-\langle \widetilde{X}^{n},[{r}^{mi}{X}_{i},{r}^{jp}{X}_{j}]\rangle-\langle \widetilde{X}^{p},[{r}^{mi}{X}^{i},{r}^{nj}{X}_{j}]\rangle
		=-r^{mi}r^{jp}f_{ij}\,^{n}-r^{mi}r^{nj}f_{ij}\,^{p},
	\end{align*}
	where by setting
	\begin{align*}
		r_{21}:=r^{ij}X_{j}\otimes X_{i}\otimes 1,\quad r_{31}:=r^{ij}X_{j}\otimes 1\otimes X_{i},\quad r_{32}:=r^{ij}1\otimes X_{j}\otimes X_{i},
	\end{align*}
	\normalfont
	in $\otimes^{3} \mathfrak{g}$, one can define
	\begin{align*}
		[r_{21},r_{31}]&=[r^{ij}X_{j}\otimes X_{i}\otimes 1, r^{lk} X_{k}\otimes 1\otimes X_{l}]=-r^{ij}r^{lk} \mathrm{ad}^{(r)}_{X_{j}}X_{k}\otimes X_{i}\otimes X_{l}\\
		&= r^{ij}r^{lk} [X_{k},X_{j}]\otimes X_{i}\otimes X_{l}=-r^{ij}r^{lk}f_{kj}\,^{s}X_{s} \otimes X_{i}\otimes X_{l},
	\end{align*}
	\begin{align*}
		[r_{21},r_{32}]&=[r^{ij}X_{j}\otimes X_{i}\otimes 1,r^{lk}1\otimes X_{k}\otimes X_{l}]= -r^{ij}r^{lk}X_{j}\otimes \mathrm{ad}^{(r)}_{X_{i}}X_{k}\otimes X_{l}\\
		&=-r^{ij}r^{lk}X_{j}\otimes [X_{k},X_{i}]\otimes X_{l}=-r^{ij}r^{lk}f_{ki}\,^{s}X_{j}\otimes X_{s}\otimes X_{l},
	\end{align*}
	\begin{align*}
		[r_{31},r_{32}]&=[r^{ij}X_{j}\otimes 1\otimes X_{i},r^{lk}1\otimes X_{k}\otimes X_{l}]=-r^{ij}r^{lk}X_{j}\otimes X_{k}\otimes \mathrm{ad}^{(r)}_{X_{i}}X_{l}\\
		&=- r^{ij}r^{lk}X_{j}\otimes X_{k}\otimes [X_{l},X_{i}]=-r^{ij}r^{lk}f_{li}\,^{s}X_{j}\otimes X_{k}\otimes X_{s},
	\end{align*}
	\normalfont
	so that,
	\begin{align*}
		[r_{21},r_{31}](\widetilde{X}^{m}\otimes \widetilde{X}^{n}\otimes \widetilde{X}^{p})
		=-r^{nj}r^{pk}f_{kj}\,^{m}
	\end{align*}
	\begin{align*}
		[r_{21},r_{32}](\widetilde{X}^{m}\otimes \widetilde{X}^{n}\otimes \widetilde{X}^{p})
		=-r^{im}r^{pk}f_{ki}\,^{n}
	\end{align*}
	\begin{align*}
		[r_{31},r_{32}](\widetilde{X}^{m}\otimes \widetilde{X}^{n}\otimes \widetilde{X}^{p})
		=-r^{im}r^{ln}f_{li}\,^{p}
	\end{align*}		
	then, we have
	\begin{align*}
		[[r,r]](\widetilde{X}^{m}\otimes \widetilde{X}^{n}\otimes \widetilde{X}^{p})=[r_{21},r_{31}](\widetilde{X}^{m}\otimes \widetilde{X}^{n}\otimes \widetilde{X}^{p})+[r_{21},r_{32}](\widetilde{X}^{m}\otimes \widetilde{X}^{n}\otimes \widetilde{X}^{p}),
	\end{align*}
	or
	\begin{align*}
		[[r,r]]=[r_{21},r_{31}]+[r_{21},r_{32}].
	\end{align*}
\end{proof}	
		
Therefore, the Yang-Baxter equation for the left Leibniz algebra can be rewritten as follows\footnote{Note that, our definitions for $r$-matrix and Yang-Baxter equations for the Leibniz algebra are different from the definition given in Ref. \cite{Raul1}.}:
\begin{align}
	[[r,r]]=[r_{21},r_{31}]+[r_{21},r_{32}]=0.\label{yang-left}
\end{align}
In the same way, we have the following proposition for the right Leibniz algebra.
\begin{prop}
	If $(\mathfrak{g},\gamma)$ be a Leibniz bialgebra such that $\mathfrak{g}$ and $\mathfrak{g}^{\ast}$ be both right Leibniz algebras, one can rewrite
	\textbf{Yang-Baxter equation} \eqref{r-bracketforleft} in terms of the structure constants  as follows:
	\begin{align}
		[[r,r]]=[r_{12},r_{13}]+[r_{12},r_{23}]=0,\label{yang-right}
	\end{align}
where 
\begin{align*}
	r_{12}:=r^{ij}X_{i}\otimes X_{j}\otimes 1,\quad r_{13}:=r^{ij}X_{i}\otimes 1\otimes X_{j},\quad r_{23}:=r^{ij}1\otimes X_{i}\otimes X_{j}.
\end{align*}
\end{prop}
\section{Examples}
\hspace{13.5pt}
In this section, we give some examples of Leibniz bialgebras, and if they are  cobundary Leibniz bialgebras then  we will compute classical $r$-matrices. For this purpose, we first rewrite the $1$-cocycle conditions  \eqref{1-cocyclelorr-r}-\eqref{1-cocyclelorr-l} in terms of the structure constants of the Leibniz algebra $\mathfrak{g}$ and $\mathfrak{g}^{\ast}$.
Using \eqref{constantg&dual}, \eqref{pairing} and \eqref{cocommutator} in the 1-cocyle conditions \eqref{1-cocyclelorr-r}-\eqref{1-cocyclelorr-l}, we obtain the following relations, respectively:
\begin{align}
&f_{ij}\,^{k}\widetilde{f}^{mn}\,_{k}=\widetilde{f}^{m'n}\,_{j}f_{im'}\,^{m}+\widetilde{f}^{m'n}\,_{i}f_{m'j}\,^{m},\label{1-cocyle1''}\\
&f_{ij}\,^{k}\widetilde{f}^{mn}\,_{k}=\widetilde{f}^{mn'}\,_{j}f_{in'}\,^{n}+\widetilde{f}^{m'n}\,_{j}f_{im'}\,^{m},\label{1-cocycle3''}\\
&f_{ij}\,^{k}\widetilde{f}^{mn}\,_{k}=\widetilde{f}^{mn'}\,_{j}f_{in'}\,^{n}+\widetilde{f}^{mn'}\,_{i}f_{n'j}\,^{n}.\label{1-cocycle4''}
\end{align}
Now, to use these relations in the calculations, we must first translate the tensor form of these relations to the matrix forms by using the following adjoint representations:
\begin{align}
f_{ij}\,^{k}=-(\chi_{i})_{j}\,^{k}=-(\mathcal{Y}^{k})_{ij}=-({(\mathcal{Y}^{k})^{t}})_{ji}=f'_{ji}\,^{k}=-(\chi'_{j})_{i}\,^{k},\label{matg}\\
\widetilde{f}^{ij}\,_{k}=-(\widetilde{\chi}^{i})^{j}\,_{k}=-(\widetilde{\mathcal{Y}_{k}})^{ij}=-(\widetilde{(\mathcal{Y}_{k})}^{t})^{ji}=
\widetilde{f}'^{ji}\,_{k}=-(\widetilde{\chi'}^{j})^{i}\,_{k}.\label{matdualg}
\end{align}
Then, the relations \eqref{1-cocyle1''}-\eqref{1-cocycle4''} have the following matrix forms, respectively:
\begin{align}
& \mathcal{Y}^{m}\widetilde{\chi'}^{n}+(\widetilde{\chi'}^{n})^{t}\mathcal{Y}^{m}
-(\widetilde{\chi}^{m})^{n}\,_{k}\mathcal{Y}^{k}=0,\label{1-cocycle1'''}\\
& \mathcal{Y}^{n}\widetilde{\chi}^{m}+\mathcal{Y}^{m}\widetilde{\chi'}^{n}-
(\widetilde{\chi}^{m})^{n}\,_{k}\mathcal{Y}^{k}=0,\label{1-cocycle3'''}\\
& \mathcal{Y}^{n}\widetilde{\chi}^{m}+(\widetilde{\chi}^{m})^{t}Y^{n}
-(\widetilde{\chi}^{m})^{n}\,_{k}\mathcal{Y}^{k}=0,\label{1-cocycle4'''}
\end{align}
where in the above relations $t$ stands for the transpose of a matrix.
On the other hand, the right and left Leibniz identities \eqref{RLI} and \eqref{LLI}  for the Leibniz algebra $\mathfrak{g}$ can be rewritten in terms of the structure constants as follows:
\begin{align*}
& f_{jk}\,^{p}f_{pi}\,^{m}=
f_{ji}\,^{p}f_{pk}\,^{m}+f_{ki}\,^{p}f_{jp}\,^{m},\\
&f_{jk}\,^{p}f_{ip}\,^{m}=f_{ij}\,^{p}f_{pk}\,^{m}+f_{ik}\,^{p}f_{jp}\,^{m}.
\end{align*}
And, similarly, we can write the following relations for the dual Leibniz algebra $\mathfrak{g}^{\ast}$:
\begin{align}
& \widetilde{f}^{jk}\,_{p}\widetilde{f}^{pi}\,_{m}
=\widetilde{f}^{ji}\,_{p}\widetilde{f}^{pk}\,_{m}
+\widetilde{f}^{ki}\,_{p}\widetilde{f}^{jp}\,_{m},\label{DRLI}\\
& \widetilde{f}^{jk}\,_{p}\widetilde{f}^{ip}\,_{m}
=\widetilde{f}^{ij}\,_{p}\widetilde{f}^{pk}\,_{m}
+\widetilde{f}^{ik}\,_{p}\widetilde{f}^{jp}\,_{m},\label{DLLI}
\end{align}
where we have the following matrix form of the relation \eqref{DRLI} and \eqref{DLLI}, respectively:
\begin{align}
(\widetilde{\chi}^{j})^{i}\,_{p}\widetilde{\chi}^{p}-\widetilde{\chi}^{j}\widetilde{\chi'}^{i}+\widetilde{\chi'}^{i}\widetilde{\chi}^{j}=0,\label{MFDRLI}\\
(\widetilde{\chi}^{i})^{j}\,_{p}\widetilde{\chi}^{p}+(\widetilde{\chi}^{i})\widetilde{\chi}^{j}-\widetilde{\chi}^{j}\widetilde{\chi}^{i}=0.\label{MFDLLI}
\end{align}
Now, one can use the relations \eqref{1-cocycle1'''}-\eqref{1-cocycle4'''} and \eqref{MFDRLI}-\eqref{MFDLLI} for the calculation of the dual Leibniz algebra $\mathfrak{g}^{\ast}$. According to this fact that $\mathfrak{g}^{\ast}$ can be left or right Leibniz algebra for the left Leibniz algebra $\mathfrak{g}$, we must solve the following equations:
\begin{itemize}
\item 
If $\mathfrak{g}$ is a left Leibniz algebra and $\mathfrak{g}^{\ast}$  is a right Leibniz algebra:	
\begin{align}
\begin{split}
\begin{cases}
 \mathcal{Y}^{m}\widetilde{\chi'}^{n}+(\widetilde{\chi'}^{n})^{t}\mathcal{Y}^{m}
-(\widetilde{\chi}^{m})^{n}\,_{k}\mathcal{Y}^{k}=0,\cr
 (\widetilde{\chi}^{j})^{i}\,_{p}\widetilde{\chi}^{p}-
\widetilde{\chi}^{j}\widetilde{\chi'}^{i}+\widetilde{\chi'}^{i}\widetilde{\chi}^{j}=0.
\label{leftorright1'''-right}
\end{cases}
\end{split}
\end{align}

\begin{align}
	\begin{split}
		\begin{cases}
			\mathcal{Y}^{n}\widetilde{\chi}^{m}+\mathcal{Y}^{m}\widetilde{\chi'}^{n}
			-(\widetilde{\chi}^{m})^{n}\,_{k}\mathcal{Y}^{k}=0,\cr
			(\widetilde{\chi}^{j})^{i}\,_{p}\widetilde{\chi}^{p}
			-\widetilde{\chi}^{j}\widetilde{\chi'}^{i}+\widetilde{\chi'}^{i}\widetilde{\chi}^{j}=0.\label{left3'''-right}
		\end{cases}
	\end{split}
\end{align}
\item 
If $\mathfrak{g}$ and $\mathfrak{g}^{\ast}$  are both  left Leibniz algebras:
\begin{align}
\begin{split}
\begin{cases}
 \mathcal{Y}^{n}\widetilde{\chi}^{m}+(\widetilde{\chi}^{m})^{t}\mathcal{Y}^{n}-
(\widetilde{\chi}^{m})^{n}\,_{k}\mathcal{Y}^{k}=0,\cr
 (\widetilde{\chi}^{i})^{j}\,_{p}\widetilde{\chi}^{p}
+(\widetilde{\chi}^{i})\widetilde{\chi}^{j}-\widetilde{\chi}^{j}\widetilde{\chi}^{i}=0.
\label{leftorright4'''-left}
\end{cases}
\end{split}
\end{align}

\begin{align}
	\begin{split}
		\begin{cases}
			\mathcal{Y}^{n}\widetilde{\chi}^{m}+\mathcal{Y}^{m}\widetilde{\chi'}^{n}
			-(\widetilde{\chi}^{m})^{n}\,_{k}\mathcal{Y}^{k}=0,\cr
			(\widetilde{\chi}^{i})^{j}\,_{p}\widetilde{\chi}^{p}
			+(\widetilde{\chi}^{i})\widetilde{\chi}^{j}-\widetilde{\chi}^{j}\widetilde{\chi}^{i}=0.\label{left3'''-left}
		\end{cases}
	\end{split}
\end{align}
\end{itemize}
Furthermore, for determining the classical $r$-matrix of a Leibniz algebra, one can rewrite the relations  \eqref{rewritecocomm1'-left} and
\eqref{rewritecocomm4'-left} in the matrix forms as follows:
\begin{align}
&\widetilde{\mathcal{Y}}_{m}=-{\chi'_{m}}^{t}r,\label{r-matrix1'left}\\
&\widetilde{\mathcal{Y}}_{m}=-r\chi'_{m}.\label{r-matrix4'left}
\end{align}
Now, using the above relations, we present some examples.
\begin{ex}
Consider the following two dimensional left Leibniz algebra \cite{Raul1}
\begin{align*}
[X_{1},X_{1}]=X_{2}\quad,\quad [X_{1},X_{2}]=X_{2},
\end{align*}
\end{ex}
By solving the  system  of equations \eqref{leftorright1'''-right} -\eqref{left3'''-left} we obtain the following $\mathfrak{g}^{\ast}$ algebras:
\begin{itemize}
\item
$\mathfrak{g}^{\ast}$ is a left Leibniz algebra
\begin{align}
[\tilde{X}^{1},\tilde{X}^{2}]=-a(\tilde{X}^{1}+\tilde{X}^{2})
,\quad [\tilde{X}^{2},\tilde{X}^{2}]=a(\tilde{X}^{1}+\tilde{X}^{2})\label{dual-l1}
\end{align}
\item
$\mathfrak{g}^{\ast}$ is a Lie algebra
\begin{align}
[\tilde{X}^{1},\tilde{X}^{2}]=-a(\tilde{X}^{1}+\tilde{X}^{2}),\quad
[\tilde{X}^{2},\tilde{X}^{1}]=a(\tilde{X}^{1}+\tilde{X}^{2})\label{dual-li1}
\end{align}
\item
$\mathfrak{g}^{\ast}$ is a right Leibniz algebra
\begin{align}
[\tilde{X}^{2},\tilde{X}^{1}]=-a(\tilde{X}^{1}+\tilde{X}^{2}),\quad
[\tilde{X}^{2},\tilde{X}^{2}]=a(\tilde{X}^{1}+\tilde{X}^{2})\label{dual-r1}
\end{align}
\end{itemize}
considering \eqref{dual-l1} as a dual Leibniz algebra  and  solving the system of equations \eqref{r-matrix4'left}, we have
\begin{align*}
r=\begin{pmatrix}
a&b\cr
-a&c
\end{pmatrix},
\end{align*}
for \eqref{dual-li1} as a dual algebra there is no $r$-matrix satisfying \eqref{r-matrix1'left} and \eqref{r-matrix4'left}.  By solving the systems of equations \eqref{r-matrix1'left}  with \eqref{dual-r1} as a dual algebra $\mathfrak{g}^{\ast}$, we obtain
\begin{align*}
r=\begin{pmatrix}
a&-a\cr
b&c
\end{pmatrix},
\end{align*}
where in the above relations $a$, $b$ and $c$ are any nonzero real numbers such that for \eqref{dual-l1} as a dual algebra, if $c=a$ and $b=-a$ then $r$ will be a classical $r$-matrix satisfying \eqref{yang-left}.
\begin{ex}\label{karbord}
Consider the following  left (and also right) two dimensional Leibniz algebra \cite{Raul1}
\begin{align*}
[X_{1},X_{1}]=X_{2}.
\end{align*}
\end{ex}
By solving the  system of equations \eqref{leftorright1'''-right}-\eqref{left3'''-left},  we have
\begin{itemize}
\item
$\mathfrak{g}^{\ast}$ is a Lie algebra
\begin{align}
[\tilde{X}^{1},\tilde{X}^{2}]=-a\tilde{X}^{1},\quad [\tilde{X}^{2},\tilde{X}^{1}]=a\tilde{X}^{1},\label{dual-l3}
\end{align}
\item
$\mathfrak{g}^{\ast}$ is a left and right Leibniz algebra
\begin{align}
[\tilde{X}^{2},\tilde{X}^{2}]=a\tilde{X}^{1},\label{dual-lr3}
\end{align}
\end{itemize}
by solving \eqref{r-matrix1'left} and \eqref{r-matrix4'left} for \eqref{dual-lr3} as a dual algebra, we have
\begin{align*}
r_1=\begin{pmatrix}
0&a\cr
b&c
\end{pmatrix},\qquad
r_2=\begin{pmatrix}
0&-a\cr
b&c
\end{pmatrix},\qquad
r_3=\begin{pmatrix}
0&b\cr
a&c
\end{pmatrix},\qquad
r_4=\begin{pmatrix}
0&b\cr
-a&c
\end{pmatrix},
\end{align*}
where $a$, $b$ and $c$ are any nonzero real numbers. If $b=-a$, then, $r_4$ and $r_1$ satisfy  \eqref{yang-left} and \eqref{yang-right}, respectively.
\begin{ex}
Consider the following three dimensional left Leibniz algebra\cite{Nil}
\begin{align*}
[X_{3},X_{1}]=X_{1}\quad,\quad [X_{3},X_{2}]=X_{2}\quad ,\quad [X_{2},X_{3}]= X_{2}.
\end{align*}
\end{ex}
Then, we have
\begin{align*}
&\chi_{1}=\begin{pmatrix}
0&0&0\cr
0&0&0\cr
0&0&0
\end{pmatrix},\qquad
\chi_{2}=\begin{pmatrix}
0&0&0\cr
0&0&0\cr
0&-1&0
\end{pmatrix},\qquad
\chi_{3}=\begin{pmatrix}
-1&0&0\cr
0&-1&0\cr
0&0&0
\end{pmatrix},\cr
&{\chi'}_{1}=\begin{pmatrix}
0&0&0\cr
0&0&0\cr
-1&0&0
\end{pmatrix},\qquad
{\chi'}_{2}=\begin{pmatrix}
0&0&0\cr
0&0&0\cr
0&-1&0
\end{pmatrix},\qquad
{\chi'}_{3}=\begin{pmatrix}
0&0&0\cr
0&-1&0\cr
0&0&0
\end{pmatrix},\cr
&\mathcal{Y}^{1}=\begin{pmatrix}
0&0&0\cr
0&0&0\cr
-1&0&0
\end{pmatrix},\qquad
\mathcal{Y}^{2}=\begin{pmatrix}
0&0&0\cr
0&0&-1\cr
0&-1&0
\end{pmatrix},\qquad
\mathcal{Y}^{3}=\begin{pmatrix}
0&0&0\cr
0&0&0\cr
0&0&0
\end{pmatrix}.
\end{align*}
Solving the systems of equations \eqref{leftorright1'''-right}-\eqref{leftorright4'''-left}, one can obtain the following $\mathfrak{g}^{\ast}$ algebras:
\begin{itemize}
\item
$\mathfrak{g}^{\ast}$ is a left Leibniz algebra
\begin{enumerate}
\item
\begin{align}
[\tilde{X}^{1},\tilde{X}^{2}]=a\tilde{X}^{2},\quad [\tilde{X}^{3},\tilde{X}^{2}]=b\tilde{X}^{2},\label{dual-l4}
\end{align}
\item
\begin{align*}
[\tilde{X}^{2},\tilde{X}^{1}]=a\tilde{X}^{1},
\quad[\tilde{X}^{3},\tilde{X}^{1}]=b\tilde{X}^{1},
\end{align*}
\item
\begin{align*}
[\tilde{X}^{3},\tilde{X}^{1}]=a\tilde{X}^{1},\quad [\tilde{X}^{3},\tilde{X}^{2}]=b\tilde{X}^{2}.
\end{align*}
\end{enumerate}
\item
$\mathfrak{g}^{\ast}$ is a right Leibniz algebra
\begin{enumerate}
\item
\begin{align}
[\tilde{X}^{1},\tilde{X}^{2}]=a\tilde{X}^{1}
,\quad [\tilde{X}^{1},\tilde{X}^{3}]=b\tilde{X}^{1},\label{dual-r4}
\end{align}
\item
\begin{align*}
[\tilde{X}^{2},\tilde{X}^{1}]=a\tilde{X}^{2},\quad [\tilde{X}^{2},\tilde{X}^{3}]=b\tilde{X}^{2}
\end{align*}
\item
\begin{align*}
[\tilde{X}^{1},\tilde{X}^{3}]=a\tilde{X}^{1},\quad [\tilde{X}^{2},\tilde{X}^{3}]=b\tilde{X}^{2},
\end{align*}
\end{enumerate}	
\end{itemize}
\section{Physical application}
In this section, we want to construct a dynamical system on Leibniz manifold using Leibniz bialgebra. Let us first review some definitions of dynamic on Leibniz manifolds \cite{ortega}. 
\begin{dfn}\cite{ortega}
Let $M$ be a smooth manifold and $C^{\infty}(M)$ be the ring of smooth functions on it. A Leibniz bracket on $M$ is a bilinear map
$\{.,.\}_{L}:C^{\infty}(M)\times C^{\infty}(M)\longrightarrow C^{\infty}(M)$ such
that  it is a derivation on each entry, i.e.
\begin{align}
\{fg,h\}_{L}=\{f,h\}_{L}g+f\{g,h\}_{L},\qquad \{f,gh\}_{L}=g\{f,h\}_{L}+h\{f,g\}_{L},
\end{align}
for any $f,g,h\in C^{\infty}(M)$. The pair $(M,\{.,.\}_{L})$ is called Leibniz manifold.
\begin{itemize}
\item
If we have
\begin{align*}
\{f,g\}_{L}=-\{g,f\}_{L},\quad \forall f,g\in C^{\infty}(M),
\end{align*}
then, the pair $(M,\{.,.\}_{L})$ will be an almost Poisson manifold.
A function $f\in C^{\infty}(M)$ with $\{f,g\}_{L}=0$ $(\{g,f\}_{L}=0)$ for any $g\in C^{\infty}(M)$ is called a left (right) Casimir of the Leibniz manifold $(M,\{.,.\}_{L})$.
\item
The Jacobiator of the bracket $\{.,.\}_{L}$ is defined as the map 
\begin{align*}
\mathcal{J}:C^{\infty}(M)\times C^{\infty}(M)\times C^{\infty}(M)\longrightarrow C^{\infty}(M),
\end{align*}
with the relation as follows
\begin{align}
\mathcal{J}(f,g,h)=\{\{f,g\}_{L},h\}_{L}+\{\{g,h\}_{L},f\}_{L}+\{\{h,f\}_{L},g\}_{L}.
\end{align} 	
A Poisson structure on $M$ is an almost Poisson structure on $M$ such that the Jacobiator is the zero map. 
\end{itemize}
\end{dfn}
\begin{rem}\cite{ortega}
Let $(M,\{.,.\}_{L})$ be a Leibniz manifold and $h$ be a smooth function 
on $M$. There exist two vector fields $X_h^R$ and $X_h^L$ on $M$ uniquely characterized by the relation:
\begin{align}
X_h^R(f)=\{f,h\}_{L},\qquad X_h^L(f)=-\{h,f\}_{L}\qquad\forall f\in C^{\infty}(M),
\end{align}
where $X_h^R$ and $X_h^L$ are called the right and left Leibniz vector field associated with the Hamiltonian function $h\in C^{\infty}(M)$, respectively. An observable $f(x)$ is a constant of motion if for the Hamiltonian $H$ we have $\{H,f\}_{L}=0$ or $\{f,H\}_{L}=0$ (more generally two observable $f_{1}$ and $f_{2}$ are in involution if $\{f_{1},f_{2}\}_{L}=\{f_{2},f_{1}\}_{L}=0$).	
\end{rem}
\begin{rem}\cite{ortega}
The flow $F_t$ of the vector field $X_h$ satisfies
\begin{align}
\frac{d}{dt}\mid_{t=0}g(F_t(m))=\{g,h\}_{L}(F_t(m)),\quad \forall g\in C^{\infty}(M)
\end{align}	
\end{rem}
\begin{rem}\cite{ortega}
Since $\{.,.\}_{L}$ and $\mathcal{J}$ are a derivation on each of their arguments, two tensor maps $B:TM^*\times TM^*\longrightarrow {\mathbb{R}}$ and $B_J:TM^*\times TM^*\times TM^*\longrightarrow {\mathbb{R}}$ can be defined as follows
\begin{align}
B(df,dg)=\{f,g\}_{L},\quad B_J(df,dg,dh)={\cal{J}}(f,g,h),\quad\forall f,g,h\in C^{\infty}(M)
\end{align}
\end{rem}
We now try to give a new method for constructing the dynamical systems  on a Leibniz manifold using the classical $r$-matrix related to coboundary Leibniz bialgebras. Let $M$ be a Leibniz manifold (as a phase space) and $x^\mu(\mu=1,2, . . ., \dim M)$ are the local coordinates of the Leibniz manifold $M$  with $2$-tensor map $B^{\mu\nu}$, then one can assign a Leibniz bracket structure
$B^{\mu\nu}$ on $M$ for arbitrary functions $f(x^\mu)$ and $g(x^\nu)$
as\footnote{If $B^{\mu\nu}$ be a combination of $G^{\mu\nu}$ (the metric of the manifold $M$) and $P^{\mu\nu}$ (the Poisson structure on $M$) then $B^{\mu\nu}$ will be a metriplectic structure \cite{Fish}.}
\begin{align}
\{f,g\}_{L}=B^{\mu\nu}\frac{\partial f}{\partial x^\mu}\frac{\partial g}{\partial x^\nu}.\label{LB}
\end{align}
In this way, one can construct dynamical variables $S_i(x^\mu),\, i=1, . . ., \dim \mathfrak{g}$ (for some Leibniz algebra $\mathfrak{g}$ as a symmetry algebra), which
are the functions on the phase space $M$, such that
\begin{align}
\{S_i,S_j\}_{L}=f_{ij}\,^kS_k,\label{findS}
\end{align}
where $f_{ij}\,^k$ is the structure constant of the Leibniz algebra 
$\mathfrak{g}$. Now, we define the right Leibniz algebra-valued functions, as
\begin{align}
Q=S_ir^{ij}X_j,\label{rightfunction}
\end{align}
and the left Leibniz algebra-valued functions, as
\begin{align}
Q=S_jr^{ij}X_i,
\end{align}
where $r=r^{ij}X_i\otimes X_j$ is the classical $r$-matrix related to the coboundary Leibniz algebra $\mathfrak{g}$. It can be shown that the following equation is the classical Yang-Baxter equation \eqref{yang-right} for the right Leibniz algebra\footnote{Note that for obtaining \eqref{l-yang-karbord}-\eqref{trace-l} one must use the relation $[X\otimes Y,Z\otimes W]=[X,Z]\otimes YW+XZ\otimes [Y,W]\,\forall X,Y,Z,W\in\mathfrak{g}$, but, this relation for the left (right) Leibniz algebra $\mathfrak{g}$, is consistent with the left (right) fundamental identity if $\mathrm{ad}^{(r)}$ and $\mathrm{ad}^{(l)}$ be both derivation, i.e., if $\mathfrak{g}$ is a left and right Leibniz algebra simultaneously. For this reason, relations \eqref{l-yang-karbord}-\eqref{trace-l} are only satisfied for especial Leibniz algebra, i.e., for (right and left) Leibniz algebras.}:
\begin{align}
\{Q\,{}_,^\otimes\, Q\}_{L}+[Q\otimes I,r]=0,\label{l-yang-karbord}
\end{align}
and, similarly, for the left Leibniz algebra  we have the following form for the classical Yang-Baxter equation \eqref{yang-left}:
\begin{align}
\{Q\, {}_, ^\otimes \,Q\}_{L}+[I\otimes Q,r]=0,\label{r-yang-karbord}
\end{align}
where in the first term of the above equations we have Leibniz bracket between $S_i$'s of the $Q$'s and the tensor product for $X_i$; in the second term we have the Leibniz algebra bracket between the tensor product of the basis of the Leibniz algebra. Now, multiplying \eqref{l-yang-karbord} by $n(Q\otimes I)^{n-1}$ and $m(I\otimes Q)^{m-1}$ from the left and right respectively; then after some calculation we have:
\begin{align}
\{\mathrm{tr}(Q^n),\mathrm{tr}(Q^m)\}+\mathrm{tr}\left([(Q\otimes I)^n,r.(I\otimes Q)^{m-1}]\right)=0.\label{trace-r}
\end{align}
In the same way, one can obtain the following equation from \eqref{r-yang-karbord}
by multiplying $n(I\otimes Q)^{n-1}$ and $m(Q\otimes I)^{m-1}$ from the left and right, respectively:
\begin{align}
\{\mathrm{tr}(Q^n),\mathrm{tr}(Q^m)\}+\mathrm{tr}\left([(I\otimes Q)^n,r.(Q\otimes I)^{m-1}]\right)=0.\label{trace-l}
\end{align}
\begin{rem}\label{repLeibniz}
Note that a simple class of Leibniz algebras can be constructed as follows
\begin{itemize}
\item
Let $V$ be a vector space and $\varphi\in V^{\ast}$ a nonzero linear functional; on ${\mathrm{Mat}}_k(V)$
indicating linear transformation (which also it is denoted by $\varphi$), $\varphi:{\mathrm{Mat}}_k(V)\longrightarrow {\mathrm{Mat}}_k(\mathbb{R})$	whose $ij$ entry is simply the original functional $\varphi$. Then, $\mathfrak{g}={\mathrm{Mat}}_k(V)$, together with the bracket 
\begin{align}
[X,Y]=\varphi(X)Y-Y\varphi(X),\label{repleft}
\end{align}
for $k\geq 2$, is a left Leibniz algebra \cite{Ongay}.
\item
Similarly, under the above assumptions $\mathfrak{g}={\mathrm{Mat}}_k(V)$, together with the bracket 
\begin{align}
[X,Y]=X\varphi(Y)-\varphi(Y)X,\label{repright}
\end{align}
is a right Leibniz algebra; for $k\geq 2$.  
\end{itemize}
For above two cases $\varphi$ is considered a homomorphism from $\mathfrak{g}={\mathrm{Mat}}_k(V)$ with Leibniz bracket $[.,.]$ to  ${\mathrm{Mat}}_k(\mathbb{R})$ with Lie bracket $[.,.]_{Lie}$
\end{rem}								
By use of remark \ref{repLeibniz}, the second terms of \eqref{trace-r} and \eqref{trace-l} are zero, one can obtain the following functions which are in involution.
\begin{align}
I_k=\mathrm{tr}(Q^k),\label{invelution}
\end{align}
Let us consider an example. Let $M=\mathfrak{g}_1\subset \mathbb{R}^3$ be the Lie algebra with the following Lie bracket
\begin{align}
[e_2,e_3]_{Lie}=e_1,
\end{align}
and assume the following Poisson structure $P$ and metric $G$ on $\mathfrak{g}_1$ \cite{Fish}
\begin{align}
P=\begin{pmatrix}
0&0&0\cr
0&0&x^1\cr
0&-x^1&0
\end{pmatrix},\quad
G=\begin{pmatrix}
0&0&0\cr
0&\alpha&\beta\cr
0&\beta&\gamma
\end{pmatrix},
\end{align}
where $\alpha$, $\beta$ and $\gamma$ are  nonzero real numbers. Then, the metriplectic structure on $\mathfrak{g}_1$ can be obtained as follows
\begin{align}
B=P+G=\begin{pmatrix}
0&0&0\cr
0&\alpha&x^1+\beta\cr
0&\beta-x^1&\gamma
\end{pmatrix}.
\end{align} 
Let the symmetry algebra $\mathfrak{g}$ be a Leibniz algebra mentioned in the example \ref{karbord}  where it is  a left and also right Leibniz algebra. Assuming the following Leibniz representation
\begin{align*}
X_1=\begin{pmatrix}
n&j\cr
k&m
\end{pmatrix},\quad
X_2=\begin{pmatrix}
e&f\cr
g&h
\end{pmatrix},\quad
\varphi(X_1)=\begin{pmatrix}
z_1&z_2\cr
z_3&z_4
\end{pmatrix},\quad
\varphi(X_2)=\begin{pmatrix}
y_1&y_2\cr
y_3&y_4
\end{pmatrix},
\end{align*}
where $n,j,k,m,e,f,g,h,z_1,z_2,z_3,z_4,y_1,y_2,y_3,y_4$ are real numbers and using \eqref{repleft}, we have
\begin{align}
X_1=\begin{pmatrix}
n&j\cr
0&m
\end{pmatrix},\quad
X_2=\begin{pmatrix}
0&(m-n)z_2\cr
0&0
\end{pmatrix},\quad
\varphi(X_1)=\begin{pmatrix}
z_1&z_2\cr
0&z_4
\end{pmatrix},\quad
\varphi(X_2)=\begin{pmatrix}
y_4&0\cr
0&y_4
\end{pmatrix}.
\end{align}
Now, one can find a solution of the equations \eqref{findS}  using \eqref{LB}, as follows	
\begin{align}
S_1(x^1,x^2,x^3)=-\frac{(-\alpha x^3+\beta x^2)\sqrt{(\alpha\gamma-\beta^2)\alpha x^1}}{(\alpha\gamma-\beta^2)\alpha}+x^1,\,\, S_2(x^1,x^2,x^3)=x^1
\end{align}
Then, from \eqref{rightfunction} one can obtain $Q$ for this example, as follows
\begin{align}
Q=\begin{pmatrix}
-anS_2&(m-n)z_2(-aS_1+cS_2)-ajS_2\cr
0&-amS_2
\end{pmatrix}.
\end{align}
Finally, from \eqref{invelution}, we have
\begin{align}
I_2=\mathrm{tr}(Q^2)=a^2(n^2+m^2)S_2^2,\qquad
I_3=\mathrm{tr}(Q^3)=-a^3(n^3+m^3)S_2^3
\end{align}
which are  the constants of motion.
\section{Conclusions and open problems}
We defined the Leibniz bialgebra, the Yang-Baxter equation and classical $r$-matrices for the Leibniz algebras. Also, we obtained a dynamical system on Leibniz manifold such that its symmetry algebra is a Leibniz algebra. There are some open problems as follows. For the quantization of Leibniz algebras one can use the Leibniz bialgebras (similar to the Lie algebras). The question is that, is the Leibniz bialgebra  an algebraic structure of the Lie rack \cite{Kin} (Leibniz-Lie rack) such that the Leibniz structure \cite{Grab} over it, is compatible with the rack structure? Also, what is the role of the Leibniz bialgebras in the integrable metriplectic systems \cite{Fish}.
\subsection*{Acknowledgment}
\vspace{3mm}
We would like to express our deepest gratitude to M. Akbari-Moghanjoughi for carefully reading the manuscript and his useful comments.
This research was supported by Azarbaijan Shahid Madani University (Research Fund No. 27.d.1518).

\end{document}